# Inkjet-printed vertically-emitting solid-state organic lasers

*O. Mhibik[1], S. Chénais[1], S. Forget[1], C. Defranoux[2], S. Sanaur[3]*

[1] *Laboratoire de Physique des Lasers – Sorbonne Paris Cité, Université PARIS 13 et CNRS (UMR 7538), 93430 Villetaneuse – France*

[2] *SEMILAB Semiconductor Physics Laboratory Co. Ltd, Hungary*

[3] *Centre Microélectronique de Provence, Ecole Nationale Supérieure des Mines de Saint Etienne, Gardanne - FRANCE*

In this paper, we show that Inkjet Printing can be successfully applied to external-cavity vertically-emitting thin-film organic lasers, and can be used to generate a diffraction-limited output beam with an output energy as high as 33.6 µJ with a slope efficiency S of 34%. Laser emission shows to be continuously tunable from 570 to 670 nm using an intracavity polymer-based Fabry-Perot etalon. High-optical quality films with several µm thicknesses are realized thanks to ink-jet printing. We introduce a new optical material where EMD6415 commercial ink constitutes the optical host matrix and exhibits a refractive index of 1.5 and an absorption coefficient of 0.66 cm$^{-1}$ at 550-680 nm. Standard laser dyes like Pyromethene 597 and Rhodamine 640 are incorporated in solution to the EMD6415 ink. Such large size "printed pixels" of 50 mm$^2$ present uniform and flat surfaces, with roughness measured as low as 1.5 nm in different locations of a 50µm x 50µm AFM scan.

Finally, as the gain capsules fabricated by Inkjet printing are simple and do not incorporate any tuning or cavity element, they are simple to make, have a negligible fabrication cost and can be used as fully disposable items. This works opens the way towards the fabrication of really low-cost tunable visible lasers with an affordable technology that has the potential to be widely disseminated.

## 1. Introduction

Organic solid-state laser devices offer the promise of building broadly tunable coherent sources in the visible spectrum, at low cost [1]. Due to a large number of available emitters (from small molecular dyes [2,3] to conjugated polymers or hybrid inorganic-organic materials [4], which all can cover a large span of wavelengths) and their ability to form clean amorphous films, a diversified variety of laser designs can be implemented, with a much higher flexibility compared to inorganics in terms of building complex structures wherein different functional materials are either mixed, or deposited on top of each other, or next to one another. In addition of the traditional technique of thermal vacuum evaporation, inexpensive solution-based processes have emerged such as dip-coating [5], doctor-blading [6], horizontal dipping [7], or spin-coating [8]. The spin-coating technique has been, up to now, the most popular and the most widely used technique to realize organic lasers out of thin films with well-controlled and uniform thicknesses. However, it also suffers from several drawbacks: firstly, it is difficult to process more than one material, unless orthogonal solvents are used for instance. Secondly, typical spin coating processes only use 2-5% of the solution dispensed onto the substrate, while the remaining is flung off into the coating bowl and wasted, increasing the fabrication cost. A promising alternative for organic laser fabrication is InkJet-Printing (IJP), which is today a widely-used technique to complete the fabrication of electronic and optical devices [9–11], including Organic Light-Emitting Diodes (OLEDs) [12], Organic Field-Effect Transistors (OFETs) [13,14], Organic Photovoltaics devices (OPVs) [15], or displays [16,17]. Lateral patterning of different active materials with a good resolution [13,18] is made simple by this direct-writing, non-contact and maskless technology.

The concept of inkjet printing of photonic components has been recently transposed to organic lasers. Inkjet printing is a key enabler for unlocking the potential of organic lasers to address spectral agility, in a device structure where the active medium is composed of an array of printed "pixels" with different materials emitting at different wavelengths. Few reports of inkjet-printed lasers have been recently published. Gardiner *et al.* have demonstrated the operation of an inkjet-printed laser based on self-organized chiral nematic liquid crystals doped by dye molecules [19,20]. Each liquid crystal droplet was made to emit at the band edge of the photonic bandgap, whose spectral position could be tuned. Pulse energies attained the µJ level with good efficiencies (up to 60%) and full coverage of the visible spectrum [20]. Another printed laser was reported by Liu *et al.*, in an organic semiconductor (F8BT) and using thermally nanoimprinted distributed feedback resonators [21]. The tuning was realized by selecting the location of the pump spot on a given DFB grating, thus setting the wavelength emission. While the DFB resonator geometry is well suited for integrated photonic sources and exhibits low laser thresholds, the output beam from a 2$^{nd}$ order DFB laser is usually poorly defined and hence difficult to collect or inject into an optical fiber without losses. Output energies harvested from these devices are also low ($<$ nJ typically). This will be especially problematic for remote spectroscopic or sensing applications, or in general for all applications that require the laser beam to propagate in free space before reaching the target, or where the source and the detection unit cannot be integrated within the same chip [22].

In this context, we developed an inkjet-printed organic laser based on a vertical external cavity, that enables high-energy ($>$30 µJ) and a high conversion efficiency ($>$30 %) in a diffraction-limited beam. Funneling laser emission into the fundamental Gaussian mode of a stable cavity is a simple yet very efficient way to obtain the highest attainable brightness for a given output power, as the brightness (defined as the output peak power P per unit area and unit solid angle, which is the relevant figure-of-merit for most laser applications) is simply given by $P/\lambda^2$ in a diffraction-limited beam ($\lambda$ is the wavelength.) In a laser-pumped VECSOL, the peak powers are in the order of kW and brightness is of the order of ~$10^{15}$ W/sr/m².

The whole fabrication process is reduced to only two steps: inkjet-printing the laser medium onto a transparent substrate (*e.g.* a simple disposable microscope slide) and placing this "capsule" between two dielectric mirrors with suited coatings to form the resonant cavity (Figure 1a). A Fabry-Perot intracavity etalon consisting of a 2-µm thick transparent polymer is then added inside the cavity to perform continuous tunability within the material gain spectrum upon rotation. Hence, while coarse tuning is realized by sending the pump spot on the desired pixel with a selected dye, fine tuning is realized by rotating the etalon. The inkjet-printed "laser capsule" is totally independent from the fine tuning element and from cavity mirrors, it has a negligible cost and can therefore be disposed after use, offering a side solution to photodegradation issues that are met when organic lasers face real-world applications.

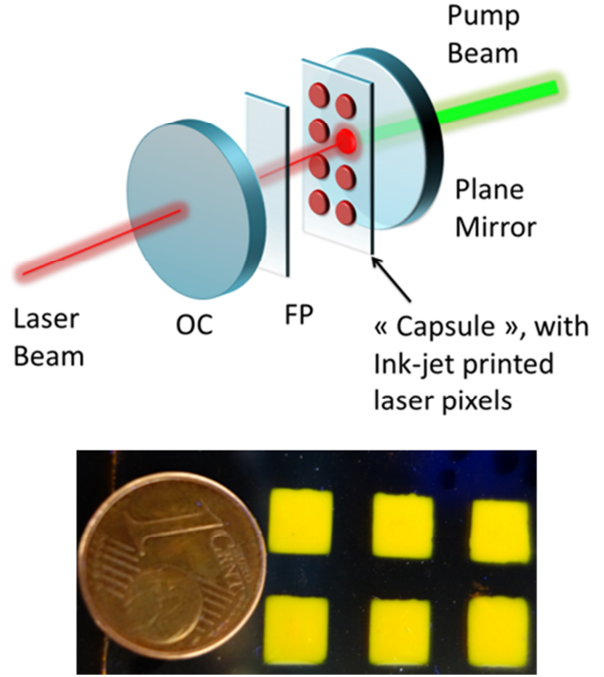

**Fig. 1**: (color online) Vertical External-Cavity Surface Emitting Organic Laser (VECSOL) setup. OC = Output Coupler, FP = Fabry-Pérot etalon (a). Pictures of inkjet printed laser pixels on transparent capsule. Here EMD6415 is incorporating Pyrromethene laser dye (b)..

The concept of the vertical external cavity surface-emitting organic laser (VECSOL) structure has been proposed previously [23,24] and a broad tunability across the visible spectrum was successfully demonstrated with spin-coated spatially uniform laser capsules. The concept was shown to be efficient to explore lasing regimes that are otherwise exotic and unreachable with typical thin-film resonators used with solution-processable gain media, such as lasing in the deep-UV thanks to nonlinear frequency conversion [25] or ultra-narrow linewidth operation ($<$ pm) in a single mode [26]. However translating the concept to an inkjet printed laser is not straightforward. In a VECSOL, the key point is to have a sufficient single-pass gain to enable the laser field to build up in a time that is significantly shorter than the pump pulse duration [27], which is challenging as the external "long" cavity ($>$ mm) is much longer than the active medium, and does not allow the laser beam to bounce many times between the mirrors. Dye-doped polymers are preferred to organic semiconductors as high fluorescence quantum yields are more easily achieved with dyes, and because dye-doped polymers enable decoupling the optical properties from the mechanical, hydrodynamical and jetting properties of the polymer host. This sets a constraint on the thickness of the film however, which has to be thick enough to enable efficient absorption of the pump light in a single pass, while keeping the dye at very low concentration to avoid luminescence quenching. In practice, using dyes with weight ratios of a few % at most is necessary, fixing the thickness to a few tens of microns [23].

## 2. Ink formulation and inkjet printing (IJP)

The choice of the matrix material to be inkjet-printed is crucial to obtain the highly transparent and smooth surface profiles that are compulsory to obtain laser action. Printing of passive optical elements has been demonstrated for years with various materials. For instance, Kumar Nallani *et al.* directly inkjet-printed microlenses of a UV cured pre-polymer [28] onto a SU8 pedestal lying on top of a VCSEL emitting facet [29]. The used pre-polymer was viscous enough to deposit 100 microns-thick microlenses but required multiple passes (300 droplets) when using a single nozzle printhead. Chen *et al.* [28] developed an in-house 100% solid (solvent free) formulation of a pre-polymer. Its viscosity could be reduced below 40 cPo at temperatures above 100°C in order to enable drop-on-demand inkjet printing [29]. Nonetheless, such UV-curing optical epoxies [28] have low thermal and chemical stability, as compared to other optical-grade plastics such as acrylics, photoresists, and thermoplastics. Such home-made ink shows optical transparency in visible spectrum. Several others papers report the fabrication of optical lenses and waveguides [30–32] by inkjet printing where the same UV cured ink attains a thickness of several microns.

Others groups inkjet-printed another type of ink for optical applications, such as the well-known SU-8 photoresist. SU-8 is an epoxy-based negative photoresist which has a high optical transparency from the visible to the near-infrared and a high refractive index (n= 1.63). Fakhfouri *et al.* used a single-nozzle printhead and applied more than hundreds of volts on the piezoelectric crystal in order to enable the jetting of SU-8 photoresist [33]. Chen *et al.* [34] diluted SU-8 photoresist to decrease its viscosity and found stable jetting for voltages of V~ 30V and pulse duration around 30 μs according the nozzle diameter. Nonetheless, the jetting was not stable for a long time. Some satellite droplets quickly appeared and spread around on the substrate. Moreover, a thickness of several microns was achieved by multiplying the number of droplets [34] on the same location.

PMMA is a reference material for photonic applications, due in particular to its excellent transparency and ability to form very low-roughness films; it was used both in previous demonstrations of VECSOLs (realized by spin coating) and for realization of passive optical waveguides [35]. We then first attempted to formulate an ink based on PMMA (solubilized in anisole). The inkjet printing technology requires some specifications. Firstly, any ink is filtered (0.2 μm pore diameter size) to remove large particle aggregates. Secondly, the printing nozzles must be kept wet, where solvents are not evaporated too fast. From a practical point of view, solvents with flash point below 110°C are not efficient. Depending on the printer we use, the viscosity of the ink has to be adjusted; here the ink should be between 10 - 12 cPo. Finally, the surface tension of the ink should be between 28 - 33 mN/m. Thus, some additives like anisole, ethylene glycol or DiMethylSulfOxide (DMSO) are necessary to adjust the volatility, the viscosity and the surface tension of the PMMA based ink. Nonetheless, the jetting reliability of such laser dye/PMMA ink is not satisfying. The waveform settings like the voltage and the pulse duration are varying for each ink preparation. Finally, some satellite droplets are formed after droplets ejection (see movie n°1 in Supplementary Information) and the jetting remains not stable enough to be satisfying.

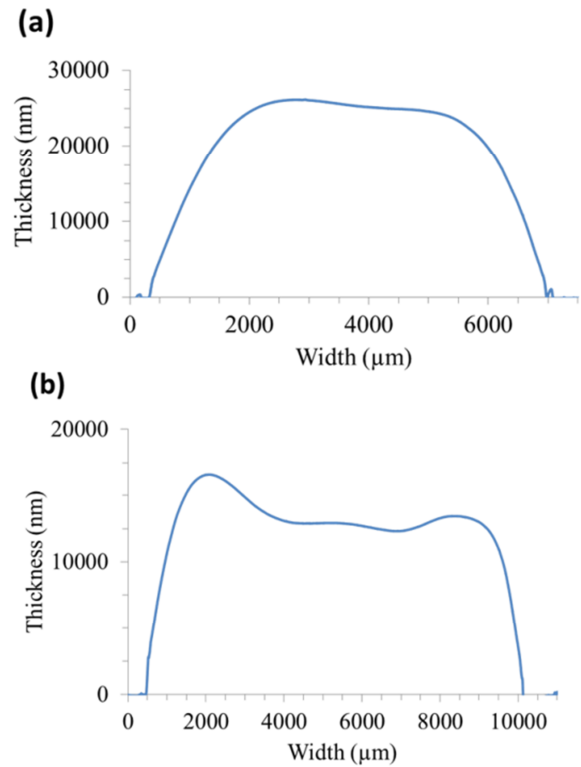

**Fig. 2** : Sample thickness and planarity according to the drop spacing of 20 μm (a) or 30 μm (b)

We then used a commercial IJP polymeric ink (SunTronic Solsys Jettable Insulator EMD 6415, Sun Chemical) which is formulated to yield a hard and nonporous surface after curing, but which had never been used to make photonic components so far to the best of our knowledge. We used efficient laser dyes (Pyrromethene 597 and Rhodamine 640, purchased from Exciton®), chosen for their high fluorescence quantum yields, good absorption in the green region and good photostability. Mixing the transparent ink with a dye will make a relevant "lasing ink" if the following criteria are fulfilled: i) good solubility of the dye into the initial ink; ii) conservation — or only minor modification — of the physical parameters that are relevant for jetting (Reynolds and Ohnesorge numbers) upon addition of the dye; iii) unchanged physical properties of the UV-cured film in presence of the dye: good wetting onto glass or silica substrate, low film roughness; iv) very high transparency of the ink outside of the absorption bands of the dye; v) ability to realize printed laser pixels of the desired thickness (around 20 µm) with a minimum number of printhead passes; vi) high flatness of the pixels over the typical size of the pump beam (~ 200 µm in diameter) to guarantee optimal beam quality.

Rhodamine 640 is solubilized in ethanol at $4\times10^{-2}$ mol/L and Pyrromethene 597 is solubilized in ethanol at $6\times10^{-2}$ mol/L. EMD 6415 ink has a viscosity of 11–13 cPo at 50°C and a surface tension of 35–37 mN.m$^{-1}$ [33]. The EMD 6415 ink is mainly based on acrylic monomers (25-40% vol.), as well as N-Vinyl caprolactam (10-25% vol.), hexamethylene diacrylate (10-25% vol.) and other acrylates for reaction of mass. EMD 6415 ink is a free-solvent ink. Polymeric layers are formed after UV curing. The lasing ink is formulated by mixing the laser dye solution (in ethanol) and the EMD 6415 ink by a ratio of 15% vol. vs 85% vol., respectively. A slight stirring is done overnight to homogenize the solution and make the dye well dispersed into the ink. The viscosity and the surface tension properties are optimized upon heating the printhead at 37°C (See movie n°2 in Supporting Information). The addition of the dye does not modify the viscosity or the surface tension of the 'lasing' ink, so that no chemical additive is needed to adjust these parameters, which validates our previous requirement ii). The printing quality depends on the drop shape, size, velocity and its interaction with the substrate, including drying conditions. The first three parameters (shape, size and velocity) are adjusted by varying the waveform parameters. The typical we use in that paper is displayed on the Figure S1. In such conditions, 30 pL droplets volume are inkjetted from the printheads, as indicated by the manufacturer. No satellite droplets are generated and all the droplets are at the same altitude after travelling 300µm distance from the printhead nozzles. This setup is mandatory to obtain faithful printed dimensions coming from the computed assisted design. Also, no particular nozzle clogging is observed, which renders the process reliable (See movie n°2 in Supporting Information). Printing conditions are also optimized by adjusting the meniscus and the drop spacing. Here, the target 20-µm thick films are obtained for an optimal drop spacing of 20µm, as seen in figure 2. In such a case, 2 layers are subsequently inkjet-printed with 15 seconds waiting between each layer printing. Then, such pixels are cured by UV exposure (Fe-Hg lamp) with a dose of 1 J/cm². The irradiance is small enough to limit the bleaching effect of dyes. To check this point, we measured the Photoluminescence Quantum Yield of the lasing ink using an integrating sphere and the experimental procedure described in [36]. Starting from an initial PLQY of 75% for the unexposed ink, we measured a slight drop to 65% for a UV dose up to 6 J/cm².

In order to fulfill our criterium iii), an $O_2$ plasma treatment was realized (110 W, 100 mTorr for 4 minutes) to reduce the dewetting effect of the lasing ink on the quartz substrates. Additionally, the quartz substrates were heated at 40°C during IJP deposition on the substrate holder in order to enhance the surface planarity.

Square pixels (approx. 50 mm²) were printed onto a quartz substrate. The Figure 1b shows a visual result of the inkjet-printed pixels. To seek the optimal drop spacing that yields the best planarity and wanted thickness, single droplets were first printed; the drop spacing was then fixed to half the diameter of a single printed droplet. For a 20 µm drop-spacing, the inkjet-printed pixel shows a very good planarity with a thickness smoothly varying from 22 to 24 µm over a 4-mm long central region (Figure 2a); in contrast, with 30 µm drop spacing (Figure 2b), the average thickness is lower (around 13µm) and the thickness fluctuations are much more pronounced. Optical microscope visualization does not show coffee-ring effects or scalloping effects on the edge of deposited droplets. The scalloped effect on the edge of a printed line is controlled by the drop spacing. If the drop spacing is too large, the scalloped effect is occurring as a narrow neck is formed between the spreading drop and the preexisting line end.

## 2. Topographical and Optical characterizations

For laser applications, roughness of the obtained films is of utmost importance as it will directly impact the amount of scattering losses in the cavity, and will have an impact on the presence of speckle on the laser beam. Measurements have been carried out with a Vecco Dimension 3100 AFM in tapping mode. An image of a typical 50µm x 50 µm sample area is shown in Figure S2. The root mean squared roughness $R_q$ is measured on 13 different locations on a given pixel and is shown to be always lower than 1.5 nm, with an averaged $R_q$ of 1 nm (Figure S3). These characteristics show a very low roughness and are comparable to what is obtained with spin-coated PMMA films.

In order to qualify the EMD6415 material as a useful material for photonics, we measured its optical properties (absorption coefficient and refractive index) after UV-curing.

The absorption spectrum of the undoped EMD6415 is measured with a visible spectrophotometer from a spin-coated film with a thickness of 30 µm (Cary Varian, Scan 100). In Figure S4, a transmission spectrum is shown for a thickness of 30 µm. The material shows excellent transparency from 400 to 800 nm, where the measured transmission is constant and equal to 99.8% between 550 and 680 nm.

The refractive index of EMD 6415 films is measured with a Semilab rotating compensator GES-5E Spectroscopic Ellipsometer at 75° of incidence. Optical modelling and a numerical regression process were carried out for all the spectra collected with Semilab Spectroscopic Ellipsometry Analyzer (SEA) software. EMD 6415 ink was deposited on silicon / silicon oxide wafer and UV cured in the same conditions than the laser capsules. Silicon oxide layer is 300 nm-thick as given as purchased. The sample is scanned from UV to near infrared wavelengths. The polarizer and the analyzer arms are tilted at a 75° angle. The amplitude ratio (⊠) and the phase shift (⊠) are fitted by a Cauchy law. This law is usually used for transparent materials in the visible region. The fit variables are the silicon oxide thickness, the EMD 6415 thickness and the Cauchy parameters applied to the EMD 6415 ink (refractive index, Cauchy coefficients). The dispersive curve of the refractive index (n) as a function of the wavelength (⊠) is shown in equation 1. The coefficient of determination ($R^2$) is equal to 0.969 and the Root Mean Square Error (RMSE) is equal to 4.06. The goodness of fit is quite reasonable and the index refractive dispersion is extracted from a Cauchy Law (equation 1).

$$n(\lambda) = 1{,}506 + \frac{0{,}00334}{\lambda^2} + \frac{0{,}00024}{\lambda^4} \qquad (1)$$

Data and fit of the amplitude ratio and the phase shift are shown in Figure S5a and S5b, respectively. The dispersion of the refractive index is shown as a function of the wavelength in Figure S5c. The refractive index ($n_0$) at long wavelengths is calculated to be around 1.5. This value is the same as others inks used for optical applications: 1.5 for home-made ink [28] and 1.57 for SU-8 photoresist or PMMA (1.5).

### 3. Laser emission

The VECSOL setup is depicted in Figure 1. The pump laser source is a frequency-doubled Q-switched diode-pumped Nd:YAG laser (Harrier from Quantronix Inc.) providing 532-nm green radiation with a 20-ns pulsewidth and a 10 Hz repetition rate. The pump beam is focused to a 200-µm-in-diameter waist onto the active capsule placed close to the input mirror and positioned at Brewster angle to remove the remaining etalon effect, which arises from the index mismatch between the dye-doped polymer film and the glass plate.

The resonator is composed of a highly reflective plane dielectric mirror (R=99.5% in the visible) and a 200-mm radius-of-curvature broadband output coupler (R=95% +/- 1% between 600 and 680 nm). The input plane mirror was highly transparent for the pump wavelength (T>95% at 532 nm).

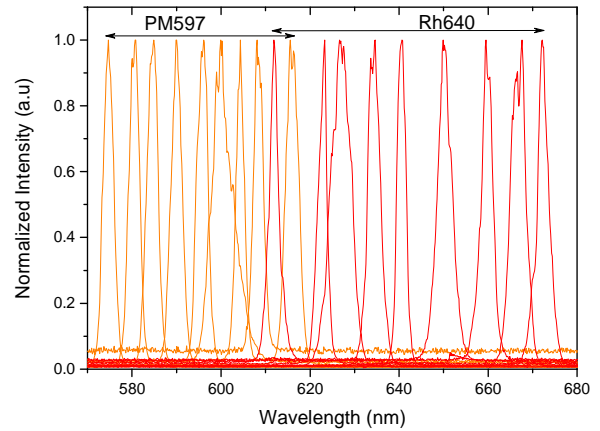

**Fig. 3** : (color online) Tunability achieved with two ink-jet printed capsules: Pyrromethene 597 (PM 597, Orange solid curves) and Rhodamine 640 (Rh 640) (red solid curves). Continuous tunability is obtained through a slight tilting of 2µm-thick intracavity freestanding polymer film.

In order to demonstrate the tuning of LASER emission and to cover a wavelength range going from yellow to deep red, two dyes belonging to the Pyrromethene and xanthene families, respectively, have been chosen for their well-established laser performance and good absorption at the pump wavelength (532 nm). Pyrromethene 597 emits in the yellow-orange region while the chosen xanthene dye (Rhodamine 640) emits in the red region. Other dyes are of course eligible to cover the blue and green part of the spectrum, providing that a UV pump is used [24]. Each dye has typically a bandwidth of 50 nm. We use a 2µm-thick freestanding film of PMMA as Fabry-Perot etalon to tune finely the laser emission. Figure 3 shows the lasing spectra recorded using a spectrometer (Jobin Yvon SPEX 270M) with a resolution of 0.8 nm. The wavelength can be continuously tuned over 40 nm for the both dye materials by tilting the etalon. Each peak has a FWHM of about 3 nm due to the weak finesse of the etalon. Figure 4 shows the output power characteristics with a 5%-transmittance-output coupler in the case of Pyromethene 597. A clear lasing threshold is visible at 2 µJ (see inset) with maximum output energy of 33.6 µJ corresponding to a slope efficiency S of 34%. Comparable results were obtained with Rhodamine 640 dye ($E_{th}$= 2µJ, S= 15%). The small-signal absorption of the ink-jet printed layer was measured to be 80% in a single pass. We compared the efficiency of the laser with the printed capsule with that of a reference laser where the gain medium consists of a dye-doped spin-coated PMMA film. The efficiency is shown to be similar as shown in Figure 4.

The laser provides a diffraction-limited output beam, as a result of the good matching between the cavity fundamental mode and the pump mode. A typical image of the $TEM_{00}$ beam profile measured at the laser output is shown in the insert of Figure 4.

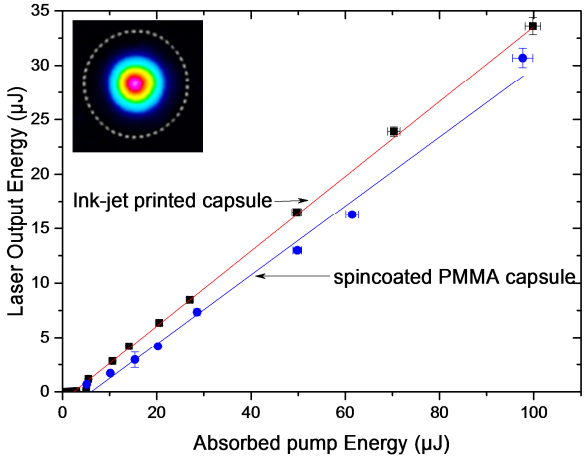

**Fig. 4** : Output laser energy versus absorbed pump energy for the Pyromethene 597 laser (Cavity length 2 mm, output coupler R=95%). The solid lines are linear fit to the experimental data. Insert: Diffraction limited beam profile measured at the exit of the laser by a CCD camera.

It is worth mentioning that the cost of a given capsule is here negligible (a few cents, corresponding to the cost of a simple microscope slide and some dye-doped ink dots), making it a disposable element, easily replaceable when degraded. The single-spot lifetime in ambient air was about $10^4$ pulses for a pump energy of 5 µJ. For a 50-mm² square printed pixel with a flat effective area of 4x4 mm² (see figure 2a), one can achieve $5.10^6$ laser pulses per pixel with a proper translation stage system. Such values are lower estimates as they correspond to conditions where the capsules are both fabricated and operated under ambient conditions, without any encapsulation. In practice also, because the laser pulse energies that are attained in this device are orders of magnitude higher than with traditional organic DFB or VCSEL lasers, a much smaller number of pulses is required to perform the same operation (treatment, sensing, etc.).

## 5. Conclusions

In this paper, we show that inkjet printing can be successfully applied to external-cavity vertically-emitting thin-film organic lasers, and can be used to generate a diffraction-limited output beam with an output energy as high as 33.6 µJ with a slope efficiency S of 34%. Laser emission show to be continuously tunable from 570 to 670 nm using an intracavity polymer-based Fabry-Perot etalon, and extension to the whole visible spectrum is straightforward with a proper choice of other dyes and UV or blue pumping. High-optical quality films with several microns thicknesses are realized thanks to inkjet printing. Indeed, EMD6415 commercial ink constitutes the optical host matrix and exhibits a refractive index of 1.5 and an absorption coefficient of 0.66 $cm^{-1}$ at 550 - 680 nm. Standard laser dyes like Pyromethene 597 and Rhodmaine 640, as used here, are incorporated in solution to the EMD6415 ink. Such large size "printed pixels" of 50 mm² present uniform and flat surfaces, with roughness measured as low as 1.5 nm in different locations of a 50µm x 50µm AFM scan. The optimal inkjet printing conditions include i) a 20µm drop spacing, ii) to heat the quartz substrates "capsules" at 40°C during printing to finally obtain "pixels" as thick as 20 µm after 2 printing passes. As the gain capsules fabricated by inkjet printing are simple and do not incorporate any tuning or cavity element, they are simple to make, have a negligible fabrication cost and can be used as fully disposable items. This work opens the way towards the fabrication of really low-cost tunable visible lasers with an affordable technology that has the potential to be widely disseminated. Such systems could find useful applications in bio and chemical analysis.

### Acknowledgements
We gratefully thank Thierry Billeton, for realizing and cutting the glass samples used as the substrates for the laser pixels, Alexis Garcia-Sanchez for realizing the AFM images, Franck Eirmbter (from SunChemical company) who provided us a sample of EMD 6415 ink, as well as DIM Nano'K Ile de France, LABEX SEAM and ANR (Vecspresso grant) for funding this work.